\documentclass[secnumarabic,aps,prd,floatfix,nofootinbib,showkeys,twocolumn,showpacs,10pt,superscriptaddress]
{revtex4-1}

\usepackage{times}
\usepackage{amsfonts}
\usepackage{amsmath}
\usepackage{amssymb}
\usepackage{natbib}
\usepackage{graphicx}
\usepackage{enumitem}
\usepackage{xcolor}
\usepackage[colorlinks=true,linkcolor=blue]{hyperref}
\usepackage[outdir=./]{epstopdf}
\usepackage[normalem]{ulem}
\usepackage{amsmath}

\def\na{New Astronomy}

\begin{document}
%%%%%%%%%%%%%%%%%%
%%%   MACROS   %%%
%%%%%%%%%%%%%%%%%%
\definecolor{orange}{rgb}{0.9,0.45,0}
\def\CovDev{D}
\def\Res{{\mathcal R}}
\def\Gammaflat{\hat \Gamma}
\def\metricflat{\hat \gamma}
\def\Dflat{\hat {\mathcal D}}
\def\part_n{\partial_\perp}
%
%=== Definition for abbreviations ===
\def\Lie{\mathcal{L}}
\def\A{\mathcal{X}}
\def\Aphi{\A_{\phi}}
\def\hAphi{\hat{\A}_{\phi}}
\def\E{\mathcal{E}}
\def\Ham{\mathcal{H}}
\def\M{\mathcal{M}}
\def\R{\mathcal{R}}
\def\p{\partial}
\def\hg{\hat{\gamma}}
\def\hA{\hat{A}}
\def\hD{\hat{D}}
\def\hE{\hat{E}}
\def\hR{\hat{R}}
\def\hcA{\hat{\mathcal{A}}}
\def\hDelt{\hat{\triangle}}
\def\na{\nabla}
\def\dif{{\rm{d}}}
\def\non{\nonumber}
\newcommand{\erf}{\textrm{erf}}
\newcommand{\saeed}[1]{\textcolor{blue}{SF: #1}} %Saeed can use command \saeed and comment in the text with his own color. 
%====================================
%
\renewcommand{\t}{\times}
\long\def\symbolfootnote[#1]#2{\begingroup%
\def\thefootnote{\fnsymbol{footnote}}\footnote[#1]{#2}\endgroup}
\title{Effect of a High-Precision Semianalytical Mass Function on the Merger Rate of Primordial Black Holes in Dark Matter Halos}

\author{Saeed Fakhry} 
\email{s\_fakhry@sbu.ac.ir}
\affiliation{Department of Physics, Shahid Beheshti University, Evin, Tehran 19839, Iran}
\affiliation{PDAT Laboratory, Department of Physics, K.N. Toosi University of Technology, P.O. Box 15875-4416, Tehran, Iran}

\author{Antonino Del Popolo}
\email{antonino.delpopolo@unict.it}
\affiliation{Dipartimento di Fisica e Astronomia, University of Catania , Viale Andrea Doria 6, 95125 Catania, Italy}
\affiliation{Institute of Astronomy, Russian Academy of Sciences, Pyatnitskaya str. 48, 119017 Moscow, Russia}

\date{\today}
\begin{abstract} 
\noindent
In this work, we study the effect of a high-precision semianalytical mass function on the merger rate of primordial black holes (PBHs) in dark matter halos. For this purpose, we first explain a theoretical framework for dark matter halo models and introduce relevant quantities such as halo density profile, concentration parameter, and a high-precision semianalytical function namely Del Popolo (DP) mass function. In the following, we calculate the merger rate of PBHs in the framework of ellipsoidal-collapse dark matter halo models while considering the DP mass function, and compare it with our previous study for the Sheth-Tormen (ST) mass function. The results show that by taking the mass of PBHs as $M_{\rm PBH}=30 M_{\odot}$, the DP mass function can potentially amplify the merger rate of PBHs. Moreover, we calculate the merger rate of PBHs for the DP mass function as a function of their mass and fraction and compare it with the black hole mergers recorded by the LIGO-Virgo detectors during the latest observing run. Our findings show that the merger rate of PBHs will fall within the LIGO-Virgo band if $f_{\rm PBH} \gtrsim \mathcal{O}(10^{-1})$. This implies that the DP mass function can be used to strengthen constraints on the fraction of PBHs.
\end{abstract}

\maketitle
\vspace{8cm}
%\tableofcontents

%%%%%%%%%%%%%%%%%%%%%%
\section{Introduction}
The study of gravitational waves (GWs), as cosmological observables, has been a focus of research for more than a few decades. Hence, in light of GWs and their detections, a new framework for evaluating astrophysical and cosmological phenomena has been provided. Binary black hole mergers are considered as one of the main sources of GWs, whose traces can potentially be found in the GW detectors \cite{Mandel:2021smh, Olsen:2022pin}. There have been dozens of binary black hole mergers detected by the Laser Interferometer Gravitational-Wave Observatory (LIGO)-Virgo collaboration in recent years, which has opened a new door in studying compact objects in the Universe \cite{Abbott:2016blz, Abbott:2016nmj, TheLIGOScientific:2016src, Abbott:2020khf, Abbott:2020tfl}. Interestingly, binary black hole mergers recorded by the LIGO-Virgo observatories have been stellar-mass black holes (those with the mass of $M_{\rm BH}< 100 M_{\odot}$). This may lead to an informative indication of the mass distribution of black holes in the Universe as well as their formation scenario. The formation channel of black holes participating in merger events recorded by the LIGO-Virgo detectors has not yet been definitively identified. Their origin may have come through stellar collapse (possibly via different channels) \cite{fishbach2021, rodriguez2021}, or they may have come through the direct collapse of primordial density fluctuations at the beginning of the Universe \cite{bird2016, Clesse:2016vqa, sasaki2016}.

It is fascinating to note that the GW data recorded by the LIGO-Virgo detectors are well consistent with the merging of primordial black holes (PBHs). PBHs are one of the potential candidates proposed to describe dark matter (see e.g., \cite{Spergel:1999mh, MACHO:2000qbb, Liebling:2012fv, Roszkowski:2017nbc, ADMX:2019uok, DiGiovanni:2020frc, DiGiovanni:2021vlu} for other dark matter candidates). The formation of PBHs is predicted to be due to the nonlinear peaks in primordial density fluctuations. Actually, primordial density fluctuations may directly collapse into PBHs when they exceed a threshold value \cite{zeldovic1967, Hawking:1971ei, Carr:1974nx}. The evolution of nonlinear density perturbations on superhorizon scales as well as the threshold amplitude of curvature perturbations for forming black holes have been considered in several studies \cite{niemar1999, shibati1999, plonarov2007, musco2013, young2014, bloomfield15, allahyari2017}. Furthermore, PBHs can span a wide range of masses, which makes them distinct from astrophysical black holes \cite{Sasaki:2018dmp}. 

As a result of their random spatial distribution, PBHs in dark matter halos might encounter each other and/or other compact objects, form binaries, emit GWs, and finally merge. In recent years, LIGO-Virgo detectors have detected GWs triggered by binary black hole mergers \cite{LIGOScientific:2018mvr, LIGOScientific:2020ibl, LIGOScientific:2021djp}, which has renewed interest in PBHs. Due to the mass range of black holes associated with binary mergers, which is often greater than the mass spectrum of astrophysical black holes, such black holes may have a primordial origin. Since PBHs are possible candidates for dark matter and can be clustered in dark matter halos, the structural characteristics of dark matter halos may affect the merger rate of PBHs. Thus, it has been indicated that some quantities like the halo mass function, halo concentration parameter, and halo density profile can improve the accuracy of theoretical models and their prediction of the merger rate of PBHs \cite{Fakhry:2020plg, Fakhry:2021tzk, Fakhry:2022uun, Fakhry:2022zum}.

Meanwhile, the halo mass function is one of the most important quantities that categorizes dark matter halos according to their mass. In other words, the halo mass function represents the mass distribution of those structures whose overdensities exceed the threshold value, separate from the expansion of the Universe, and gravitationally collapse. Therefore, in addition to describing the nature of dark energy \cite{2011PhRvL.107b1302S, Yang:2014okp, Huterer:2017buf, DiValentino:2019ffd, Y:2021ybx, Yengejeh:2022tpa}, one of the main challenges facing any cosmological model is to provide a high-precision mass function for the dark matter distribution. Using a spherical-collapse model, Press and Schechter (PS) proposed an analytical model with primordial fluctuations that evolve from the linear phase to the collapse time \cite{ps1974}. The PS mass function has been proposed as the simplest model, and it is consistent with observations in many cases. However, at some mass limits, it quantitatively differs from the numerical results. Consequently, some improvements have been made to deal with this challenge. Regarding this, one of the best formalisms was provided by Sheth and Tormen (ST) based on a more realistic model, which fits the simulation results more accurately \cite{Sheth:1999mn, Sheth:1999su, Sheth:2002}. The ST formalism was based on an ellipsoidal-collapse model with a dynamical threshold value of overdensities. However, semianalytical approaches indicate that the ST mass function also overpredicts the number of halos at large masses \cite{Reed:2003sq}. Moreover, the study of the evolution of the PS and ST mass functions reveals a worsening of the situation. It should be noted that the mass functions derived from simulations lead to convenient approximations in some cases. However, the black-box nature of simulations, which includes many physical effects, makes it hard to decipher how these mechanisms affect the shape of the halo mass function. Therefore, a suitable semianalytical approach is needed to separately determine the effect of physical processes on the shape of the halo mass function.

In this work, we investigate the effect of a high-precision semianalytical mass function on the merger rate of PBHs in dark matter halos. In this respect, the outline of the work is as follows. In Sec.~\ref{sec:iii}, we briefly discuss an appropriate dark matter halo model and some key quantities such as the halo density profile, concentration parameter, and a high-precision semianalytical function known as Del Popolo (DP) mass function. Also, in Sec.~\ref{sec:iii} we calculate the merger rate of PBHs in the ellipsoidal-collapse model while considering the DP mass function, and compare it with that obtained from the ST mass function. In addition, we compare the redshift evolution of the merger rate of PBHs obtained from the DP mass function with the corresponding results derived from the ST mass function. We also perform our analysis in terms of various PBH masses and discuss their constraints arising from the calculations with the DP mass function. Finally, in Sec.~\ref{sec:iv}, we discuss the results and summarize the findings.

%%%%%%%%%%%%%%%%%%%%%%
\section{Dark matter halo models}\label{sec:iii} 
\subsection{Halo density profile}
In the standard model of cosmology, dark matter halos are considered as nonlinear cosmological structures that have been distributed in the Universe based on the formation and evolution of hierarchical structures.  The primordial density fluctuations could have exceeded a critical value, collapsed due to self-gravitational forces, and become qualified to form dark matter halos. From a physical point of view, such conditions can be described by a dimensionless quantity known as density contrast, which is derived from the excursion sets theory. This quantity is defined as $\delta(r)\equiv (\rho(r)-\bar{\rho})/\bar{\rho}$, where $\rho(r)$ represents the density of the overdense region at an arbitrary point $r$, and $\bar{\rho}$ denotes the mean density of the background.

On the other hand, cosmological and structure formation models are constrained by the inner region of dark matter halos, whose masses can be described by a radius-dependent function called the density profile. However, as a reliable criterion, spectroscopic observations of gravitational lensing, x-ray temperature maps and stellar dynamics in galaxies can predict dark matter distribution in the central regions of galactic halos \cite{Reed:2003hp}. In recent decades, analytical approaches and numerical simulations have been used to obtain a suitable density profile in such a way that its predictions are consistent with the relevant observational data \cite{ein,Jaffe:1983iv,deZeeuw:1985sk,Hernquist:1990be,dehnen,Navarro:1995iw}. Based on $N$-body simulations in the framework of cold dark matter (CDM) models, Navarro, Frenk, and White (NFW) presented the following density profile \cite{Navarro:1995iw}
\begin{equation}\label{nfw}
\rho(r)=\dfrac{\rho_{\rm s}}{r/r_{\rm s}(1+r/r_{\rm s})^{2}},
\end{equation}
where $\rho_{\rm s}=\rho_{\rm crit}\delta_{\rm c}$ is the scaled density of the halo, $\rho_{\rm crit}$ is the critical density of the background Universe at a given redshift, and $r_{\rm s}$ is the scale radius of the halo.

On the other hand, through analytical approaches, Einasto found another suitable definition of the density profile as follows \cite{ein}
\begin{equation}\label{einasto}
\rho(r)=\rho_{\rm s}\exp\{-\dfrac{2}{\alpha}\left[\left(\dfrac{r}{r_{\rm s}}\right)^{\alpha}-1\right]\},
\end{equation}
where $\alpha$ is the shape parameter. The strength of the presented density profiles is their consistent predictions with the distribution of dark matter in galactic halos and their rotation curves. 

In addition to the density profile, there is another quantity to describe the central density of galactic halos, called the concentration parameter, which is defined as follows
\begin{equation}
C\equiv\dfrac{r_{\rm vir}}{r_{\rm s}},
\end{equation}
where $r_{\rm vir}$ is the halo virial radius. The halo virial radius covers a volume within which the average halo density is $200$ to $500$ times the critical density of the Universe. Numerical simulations and analytical approaches imply that to provide a suitable prediction, the concentration parameter must evolve dynamically with mass and redshift. Many studies have been conducted to obtain an appropriate concentration parameter \cite{prada,dutton, okoli, ludlow}. In this regard, one of the successful analytical approaches to obtain the appropriate mass-concentration-redshift relation is presented in Ref. \cite{okoli}, which is structured based on a triaxial collapse.
%%%%%%%%%%%%%%%%%%%%%%%%%%%%%%%%%%%%%%%%%%%%%%
\subsection{A high-precision mass function}
The standard model of cosmology, also known as the $\Lambda$CDM model, is successful in fitting a variety of data from intermediate to large scales \citep{WMAP:2003elm, DelPopolo:2007dna, 2011ApJS..192...18K, delpopolo2013, 2013ApJ...779...86S, Das:2013zf, DelPopolo:2013qba, Planck:2015fie}. 

The accurate prediction of the halo mass function\footnote{The mass function represents the dark matter halos mass distribution, or in other terms the number density of dark matter halos per mass interval, see \cite{DelPopolo:2007dna, Hiotelis:2005jh, Hiotelisapp}.} is another important test of the $\Lambda$CDM model. The mass function is used to determine the cosmological parameters and is also of great importance in studying the dark matter distribution and the formation and evolution of galaxies. Then a simple and accurate high-precision mass function, that can be applied to different cosmologies and redshifts is a helpful and valuable asset.

In the simple mass function model proposed by PS, the initial fluctuations are spherical and have a Gaussian distribution \cite{ps1974}. The fluctuation evolution is followed from the linear phase to the collapse by means of a spherical-collapse model. At the epoch of virialization, the density contrast, $\delta(r)$, obtained by means of the linear perturbation theory, has the value $\delta_{\rm c}\simeq 1.686$ in the case of an Einstein-de Sitter cosmology. For a density field having a Gaussian probability distribution, the probability that on a given scale the overdensity exceeds $\delta_{\rm c}$ is proportional to the number of cosmic structures having a density perturbation greater than $\delta_{\rm c}$.

Unfortunately, the PS mass function suffers from serious problems. In this regard, the number of objects in the high mass tail is underpredicted, and the reverse for the low mass tail of the mass function, see, e.g., \cite{Efstathiou:1988tk, Gross:1997wv, Jenkins:2000bv, White:2002at}. An improvement in the PS mass function can be obtained by moving from a spherical to an ellipsoidal collapse \cite{Sheth:1999su}. While the mass function obtained with the ellipsoidal collapse is in agreement with ST $N$-body simulations \citep{Sheth:1999mn, Sheth:1999su}, a deeper analysis of semianalytical models showed that the ST mass function overpredicts the halo number at large masses \citep{Lukic:2007fc}. More in detail the ST mass function gives an overestimation up to a factor of $\simeq 3$ for the rarest dark matter halos \citep{Reed:2006rw, Lukic:2007fc}. Comparing with the Bolshoi simulation \citep{2011ApJ...740..102K}, at $z=0$ the discrepancy is smaller than $10\%$ in the mass range $(10^{9}-10^{14})M_{\odot}$. Unfortunately, the discrepancy increases with redshift and at $z=10$ the ST mass function gives about $10$ times more halos than simulations.

The extended PS approach, also known as ``excursion set formalism" \citep{1991ApJ...379..440B, 1993MNRAS.262..627L}, often used to model the halo formation statistics, is based on stochastic processes. In the quoted formalism the statistics of halos start the statistical properties of the average overdensity within a window of radius $R_{\rm f}$, namely $\overline{\delta}(R_{\rm f})$. A Gaussian density field smoothed with a filter represents the density perturbations. $\overline{\delta}(R_{\rm f})$ vs $R_{\rm f}$, in a hierarchical Universe is a random walk \cite{DelPopolo:2006gn}. When the random walk crosses a threshold value, or barrier, $\delta_{\rm c}$, a dark matter halo forms. Instead of the filtering radius, sometimes other quantities are often used, e.g., the mass variance $S$. The first-crossing distribution is the probability that a random walk first crosses the threshold, also dubbed ``barrier", between $S$ and $S+dS$. The distribution of the first crossings defines the ``multiplicity function", which is connected to the mass function.

The first improvement on the PS threshold was obtained in Ref.~\cite{Popolo:1998fz}, finding that the collapse threshold is not constant as in the PS model but becomes mass dependent. It is given by
\begin{equation} \label{eqn:barrier}
\delta_{\rm cm} = \delta _{\rm c}(z) \left(1+\frac{\beta}{\nu^\alpha}\right),
\end{equation}
where $\alpha=0.585$, and $\beta=0.46$. Also, $\nu=(\delta_{\rm c}/\sigma)^{2}$, where $\sigma(M, z)$ is the linear root-mean-square fluctuation of overdensities on a comoving scale including a mass $M$ at redshift $z$.

In Ref.~\cite{Sheth:1999su}, the threshold is obtained by considering an ellipsoidal collapse as follows
\begin{equation}\label{eqn:barrier1}
\delta_{\rm ec} = \delta_{\rm c}(z) \left( 1+\frac{\beta_1}{\nu^{\alpha_1}} \right),
\end{equation}
where $\alpha_1=0.615$ and $\beta_1=0.485$. 

Based on the excursion set formalism, it can be illustrated that Eqs.~\eqref{eqn:barrier} and \eqref{eqn:barrier1} correspond to the following barriers, respectively
\begin{equation}\label{eqn:barrierv}
B(M)=\sqrt{a} \delta _{\rm c}(z) \left(1+\frac{\beta}{a \nu^\alpha} \right)\;,
\end{equation} 
and 
\begin{equation}\label{eqn:barrierv1}
B(M)_{\rm ST}=\sqrt{a_1} \delta _{\rm c}(z) \left(1+\frac{\beta_1}{a_1 \nu^{\alpha_1}}\right)\;.
\end{equation}
It should be noted that the quoted barriers with an accurate choice of $a$, and $a_1$ give a mass function in agreement with simulations, see \cite{Sheth:1999su, DelPopolo:2006bk}.

To take into account several physical effects, like fragmentation and mergers \citep{Sheth:1999su}, tidal torques \citep{Popolo:1998fz},
dynamical friction, and the cosmological constant \citep{DelPopolo:2006bk}, it is necessary to change the shape of the barrier and add in it the quoted effects. As shown in \cite{DelPopolo:2006shv}, an improved threshold, which is directly proportional through a constant to the barrier is given by
\begin{align}\label{eqn:barrierf}
\delta_{\rm cm2} & = \delta_{\rm co}\left[1+\int_{r_{\rm i}}^{r_{\rm ta}} \frac{r_{\rm ta} L^2 \cdot {\rm d}r}{G M^3 r^3} + \frac{\lambda_{\rm o}}{1-\mu(\delta)}+\Lambda \frac{r_{\rm ta} r^2}{6 G M}\right] \nonumber \\
& \simeq \delta_{\rm co} \left[1+\frac{\beta}{\nu^{\alpha}}+\frac{\Omega_{\Lambda}\beta_2}{\nu^{\alpha_2}}+\frac{\beta_3} {\nu^{\alpha_3}}\right]\;.
\end{align}
In the above equation, $\lambda_{\rm o}=\epsilon_{\rm o} T_{\rm co}$, where $T_{\rm co}$ is the collapse time of a perturbation without the effect of dynamical friction (see Appendix~A of \cite{DelPopolo:2017snx} and Eq.~(24) of \cite{1994ApJ...427...72A}). Also, $\epsilon_o$ is proportional to the dynamical friction coefficient $\eta$ (see Appendix~A of \cite{DelPopolo:2017snx} and Eq.~(23) of \cite{1994ApJ...427...72A}). Moreover, $\mu(\delta)$ is given in Eq.~(29) of \cite{Colafrancesco:1994ne} and Appendix~A of \cite{DelPopolo:2017snx}, $\Lambda$ is the cosmological constant, and $r_{\rm ta}$ is the turn-around radius. The angular momentum $L$ is calculated as shown in \cite{DelPopolo:2006shv, 2009ApJ...698.2093D, 2009A&A...502..733D} and in Appendix~ A of \cite{DelPopolo:2017snx}. In addition, the existing constants take values as $\alpha=0.585$, $\beta=0.46$, $\alpha_2=0.4$, $\beta_2=0.02$, $\alpha_3=0.45$, and $\beta_3=0.29$.

The mass function in the excursion set formalism is defined as the comoving number density of halos in a mass range $(M\mbox{-}M+dM)$
\begin{equation}\label{eqn:universal}
n(M,z)=\frac{\overline{\rho}}{M^{2}}\left|\frac{d\log{\nu }}{d\log M}\right|\nu f(\nu)\;,
\end{equation}
where $\overline{\rho}$ is the background density. The distribution of the first crossing, namely the ``multiplicity function" is indicated with $f(\nu)$. 

As shown in Ref.~\cite{DelPopolo:2017snx}, using the barrier given by Eq.~\eqref{eqn:barrierf}, taking into account the effects of angular momentum, dynamical friction, and cosmological constant one can obtain the following relation
\begin{equation}\label{nufnu}
\nu f(\nu) \simeq A_{2}\sqrt{\frac{a\nu}{2\pi}}l(\nu)\exp{\left\{-0.4019 a \nu^{2.12} m(\nu)^{2}\right\}},
\end{equation}
where
\begin{equation*}
l(\nu)=\left(1+\frac{0.1218}{\left(a\nu\right)^{0.585}}+\frac{0.0079}{\left(a\nu\right)^{0.4}}+\frac{0.1}{\left(a\nu\right)^{0.45}}\right),
\end{equation*}
and
\begin{equation*}
m(\nu)=\left(1+\frac{0.5526}{\left( a\nu \right)^{0.585}} +\frac{0.02}{\left(a\nu\right)^{0.4}}+\frac{0.07}{\left(a\nu\right)^{0.45}}\right).
\end{equation*}
Also, relevant constants take values as $A_2=0.93702$, and $a=0.707$. By incorporating Eq. \eqref{nufnu} into Eq. \eqref{eqn:universal}, one can obtain a high-precision semianalytical mass function, which we call the Del Popolo (DP) mass function from now on.

As mentioned earlier, the DP mass function is considered to be a high-precision function to describe the mass distribution of dark matter halos. In the following, the relevant arguments for this claim will be discussed. It is insightful to mention that, a high-precision analytical mass function requires a precise characterization of the barrier, whose shape depends on the likely physical effects. Naturally, under such reasoning, one would expect that the PS mass function with an oversimplified barrier provides a bad fit to the simulations. In Ref.~\cite{DelPopolo:2017snx}, a relative comparison between different mass functions \cite{Jenkins:2000bv, Sheth:2002, Reed:2003sq, 2006ApJ...646..881W, 2007MNRAS.374....2R, 2010MNRAS.403.1353C, 2011MNRAS.410.1911C, 2011ApJ...732..122B, 2012MNRAS.426.2046A} with the DP mass function has been performed. This comparison demonstrates that the DP mass function at $z=0$ is consistent on average to about $3\%$ level with all other mass functions in the range $-0.55<\log \sigma ^{-1}<1.31$, while the corresponding deviation for other mass functions (e.g., \cite{2011MNRAS.410.1911C, 2012MNRAS.426.2046A}) is greater than $3\%$. Furthermore, the redshift evolution of the ratio of the mass function obtained in \cite{2011ApJ...732..122B} to the DP mass function exhibits a discrepancy of less than $3\%$. On the other hand, in $z=0$, the ST mass function deviates by about $10\%$ compared to the simulations, and this deviation changes directly with redshift. In other words, in $z=6$, the ST mass function predicts the number of halos by $1.5$ times and in $z=10$ by $10$ times more than the simulations \cite{DelPopolo:2017snx}. Meanwhile, the DP mass function deviates by less than $3\%$ compared to the Bolshoi simulations \cite{2011ApJ...740..102K} at different redshifts. In this comparison, low-mass objects indicate relatively small deviations while high-mass ends illustrate larger deviations. Accordingly, it can be inferred that the DP mass function is capable of providing a very good prediction of the dark matter distribution in galactic halos.
%%%%%%%%%%%%%%%%%%%%%%%%%%%%%%%%%%%%%%%%%%%%%%
\section{Primordial black hole merger rate} \label{sec:iii}
As mentioned earlier, PBHs are attributed to those black holes that could have formed in the earliest stages of the Universe from the direct collapse of density fluctuations. Given that PBHs have been randomly distributed in the Universe, it is expected that they can encounter each other and form binaries. 

The first gravitational wave emitted by two merging black holes with similar masses strengthened the possibility that PBHs could form dark matter. Accordingly, a Poissonian isocurvature density fluctuation component is induced by PBHs into the power spectrum of the adiabatic density fluctuations produced through the inflationary era \cite{Afshordi:2003zb}. The isocurvature component arising from PBHs dominates the small scales, leading to significant modifications in the collapse history, such that it can initially yield much greater rates of collapse \cite{Kashlinsky:2016sdv}. Consequently, one can expect that the initial power spectrum of isocurvature Poissonian fluctuations induced by PBHs on dark matter is as follows \cite{Afshordi:2003zb}
\begin{equation}
\mathcal{P}_{\rm PBH, i}=n_{\rm PBH}^{-1},
\end{equation}
where $n_{\rm PBH}$ is the comoving mean density of PBHs. From the formation time of PBHs to the present day, the isocurvature fluctuations in subhorizon scales grow by a factor of $3/2(1+z_{\rm eq})$, where $z_{\rm eq}\simeq4000$ is the redshift corresponding to the matter-radiation equality. Therefore, the extra power spectrum at redshift $z$ is specified by \cite{Afshordi:2003zb}
\begin{equation}
\mathcal{P}_{\rm PBH}(z)=\dfrac{9}{4}(1+z_{\rm eq})^{2}\,n_{\rm PBH}^{-1} \,[D(z)]^{-2},
\end{equation}
where $D(z)$ is the linear growth factor of density fluctuations from redshift $z$ to the present day. Thus, the total power spectrum on scale $k$ and at redshift $z$ takes the following form
\begin{equation}
\mathcal{P}_{\rm tot}(k, z)=\mathcal{P}_{\rm ad}(k, z)+\mathcal{P}_{\rm PBH}(z),
\end{equation}
where $\mathcal{P}_{\rm ad}(k, z)$ is the adiabatic power spectrum of dark matter. Therefore, the physical effect of the isocurvature power spectrum caused by PBHs can be evaluated in the linear root-mean-square fluctuation of overdensities, which has the following form
\begin{equation}
\sigma^{2}(M, z)=\dfrac{1}{2\pi^{2}}\int\mathcal{P}_{\rm tot}(k, z)W(k)k^{2}dk,
\end{equation}
where $W(k)$ is the top-hat function that depends on the mass $M$ contained in wavelength $2\pi/k$. Note that in this work, we employ the total power spectrum $\mathcal{P}_{\rm tot}(k, z)$ in our calculations.

The merger time of PBH binaries that are supposed to form in dark matter halos relies on the velocity dispersion of the halos (from hours to kiloyears) \cite{OLeary:2008myb}. Accordingly, PBH binaries formed via dissipative two-body encounters have a merger time much shorter than the age of the Universe. However, nondissipative three-body encounters can also lead to the formation of PBH binaries in dark matter halos. The binaries formed through such a channel often do not have sufficient binding energy for the instantaneous emission of gravitational radiation. Therefore, nondissipative three-body encounters often yield wide binaries whose merger time is longer than a Hubble time \cite{Quinlan1989}, and naturally, they should not significantly contribute to the population of BH mergers recorded by the LIGO-Virgo detectors. However, the Poisson effect implies that PBHs can lead to a possible enhancement of small-scale structures if they can make a substantial contribution to dark matter \cite{Carr:2018rid}. Under these conditions, it is argued that three-body encounters occurring at high redshifts can make an enormous contribution to the population of BH mergers associated with the LIGO-Vigo observations, in such a way that they are comparable to that of two-body dynamical captures \cite{Korol:2019jud, Trashorras:2020mwn, Franciolini:2022ewd}.

Assume two PBHs with masses $m_{1}$ and $m_{2}$, and relative velocity $v_{\rm rel}$ at a large distance can encounter each other in dark matter halos.  Consequently, two-body scattering suggests that notable gravitational radiation propagates at the closest physical separation called the periastron \cite{peters}. PBHs will become gravitationally bound and form binary if the emitted gravitational energy exceeds the kinetic energy of the system. Hence, periastron has a maximum value as a result of this condition \cite{peters}
\begin{equation}\label{priastron}
r_{\rm mp} = \left[\dfrac{85\pi \sqrt{2} G^{7/2}m_{1}m_{2}(m_{1}+m_{2})^{3/2}}{12c^{5}v_{\rm rel}^{2}}\right]^{2/7},
\end{equation}
where $G$ is the gravitational constant and $c$ is the velocity of light. In addition, the Newtonian limit implies the following relation between the impact parameter and the periastron \cite{Sasaki:2018dmp}
\begin{equation}
b^{2}(r_{\rm p})=\dfrac{2G(m_{1}+m_{2})r_{\rm p}}{v_{\rm rel}^{2}}+r_{\rm p}^{2}.
\end{equation}
Additionally, the tidal forces produced by surrounding compact objects on the binary can be neglected once $r_{\rm p}\ll b$ is established as strong limits to gravitational focusing. As a result, the cross-section for the binary formation can be derived as follows \cite{Quinlan1989, Mouri2002}
\begin{equation}\label{crossec}
\xi(m_{1}, m_{2}, v_{\rm rel})=\pi b^{2}(r_{\rm p, max})\simeq \dfrac{2\pi G(m_{1}+m_{2})r_{\rm p, max}}{v_{\rm rel}^{2}}.
\end{equation} 
Consequently, by inserting Eq.~\eqref{priastron} into Eq.~\eqref{crossec}, one can obtain an explicit form of the cross-section for the binary formation as
\begin{equation}
\xi \simeq 2\pi \left(\dfrac{85\pi}{6\sqrt{2}}\right)^{2/7}\dfrac{G^{2}(m_{1}+m_{2})^{10/7}(m_{1}m_{2})^{2/7}}{c^{10/7}v_{\rm rel}^{18/7}}.
\end{equation}
In this work, the events of interest are those consistent with the LIGO-Virgo sensitivity. Hence, we restrict our analysis to the case where $m_{1}=m_{2}=M_{\rm PBH}$ is satisfied. In this regard, the binary formation rate within a galactic halo is given by the following formula \cite{bird2016}
\begin{equation}
\Gamma(M_{\rm vir})=\int_{0}^{r_{\rm vir}}2\pi r^{2}\left(\dfrac{f_{\rm PBH}\rho_{\rm Halo}}{M_{\rm PBH}}\right)\langle\xi v_{\rm rel}\rangle dr,
\end{equation} 
where $0 < f_{\rm PBH} \leq 1$ is the fraction of PBHs that specifies their contribution to dark matter, $\rho_{\rm halo}$ represents the halo density profile, and the angle bracket is an average over the PBH relative velocity distribution in the galactic halo.

\begin{figure}[t!]
\begin{minipage}{1\linewidth}
\includegraphics[width=0.9\textwidth]{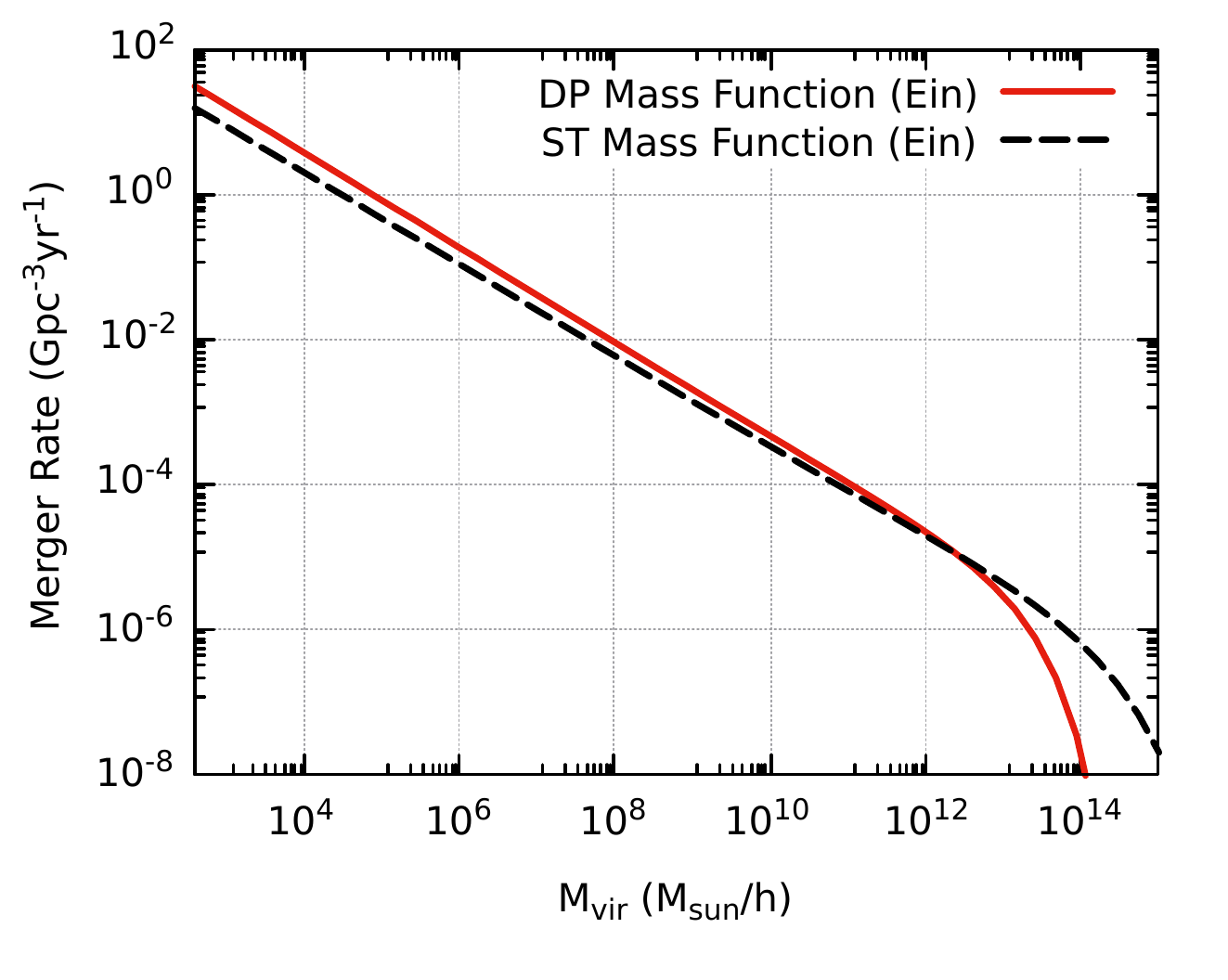}\\
\includegraphics[width=0.9\textwidth]{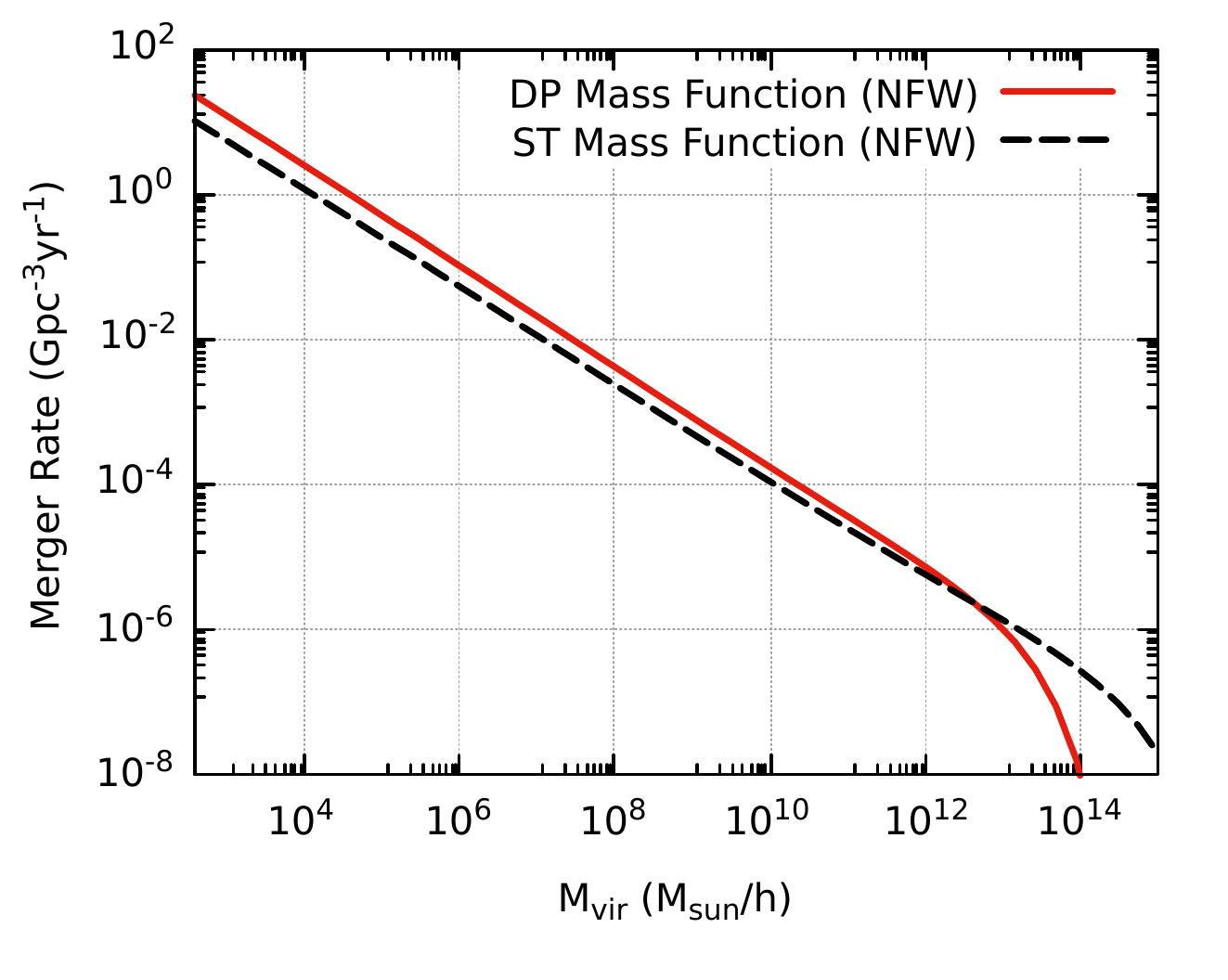}\\
\caption{Merger rate of PBHs per unit volume and time as a function of halo mass for Einasto (top) and NFW (bottom) density profiles. Solid (red) lines show this relation for DP mass function, while dashed (black) lines indicate it for ST mass function.}
\label{fig1}
\end{minipage}
\end{figure}

\begin{figure}[t!]
\begin{minipage}{1\linewidth}
\includegraphics[width=0.9\textwidth]{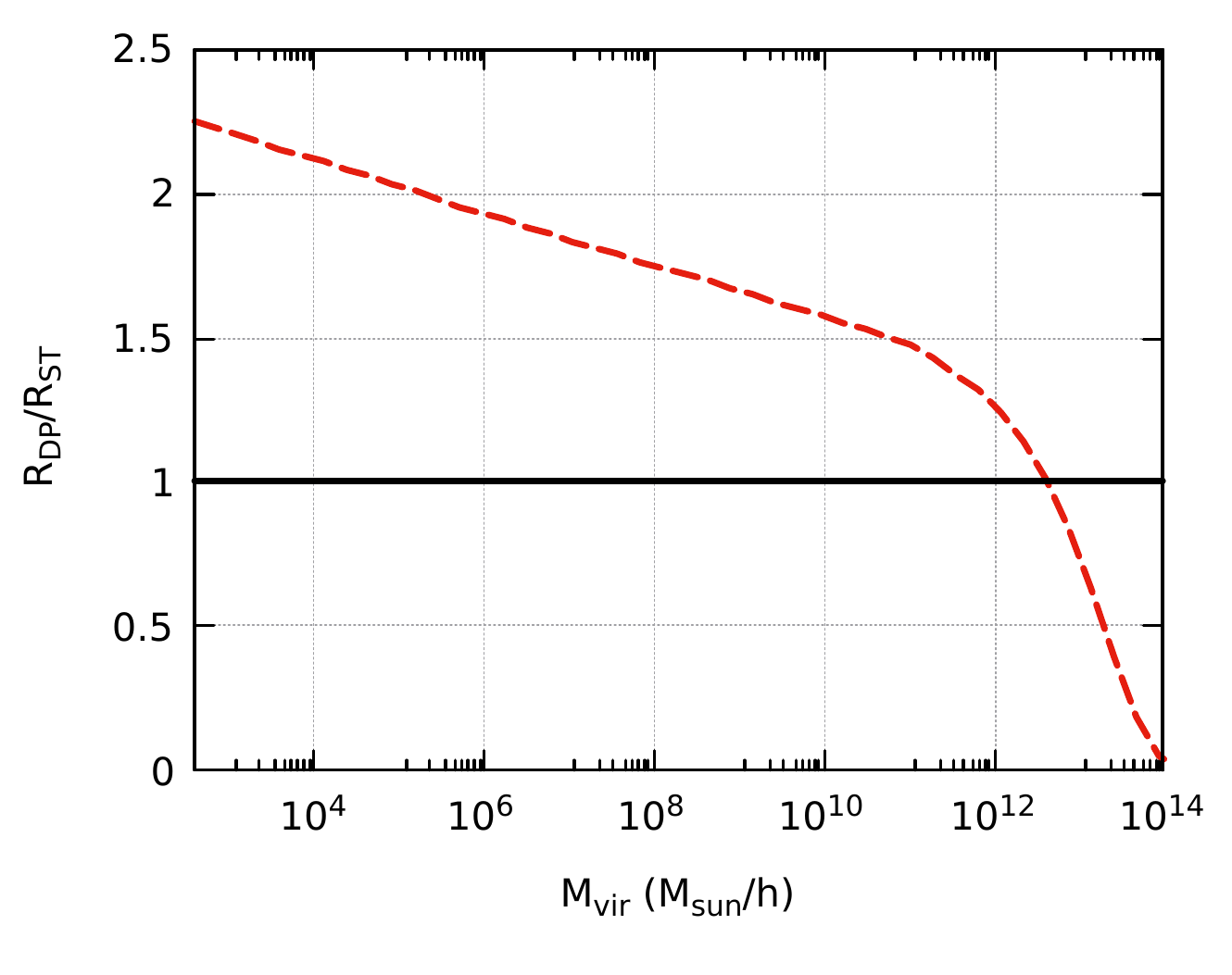}
\caption{The ratio of the merger rate of PBHs derived from the DP mass function to that obtained from the ST mass function in terms of halo mass.}
\label{fig2}
\end{minipage}
\end{figure}

Based on Ref.~\cite{prada2012}, one can calculate the halo velocity dispersion by using its relation to the maximum velocity at $r_{\rm max}$ radius
\begin{equation}
v_{\rm disp}=\dfrac{v_{\rm max}}{\sqrt{2}}=\sqrt{\dfrac{GM(r<r_{\rm max})}{r_{\rm max}}}.
\end{equation}

Assuming a cutoff at the halo virial velocity and $v_{\rm rel}=v_{\rm PBH}$, we demand that the relative velocity distribution of PBHs in the galactic halo corresponds to the Maxwell-Boltzmann statistics:
\begin{equation}
P(v_{\rm PBH}, v_{\rm disp})=A_{0}\left[\exp\left(-\dfrac{v_{\rm PBH}^{2}}{v_{\rm disp}^{2}}\right)-\exp\left(-\dfrac{v_{\rm vir}^{2}}{v_{\rm disp}^{2}}\right)\right],
\end{equation}
where $A_{0}$ is specified when the condition $4\pi \int_{0}^{v_{\rm vir}}P(v)v^{2}dv = 1$ is satisfied. 

There are two distinct mechanisms for the PBH binary formation. We mainly focus on the PBH binaries formed in dark matter halos in the late-time Universe. However, the initial clustering may lead to the separation of PBHs from the Hubble flow and yield the formation of PBH binaries in the early Universe \cite{2017PhRvD..96l3523A, 2020PhRvD.101d3015V}. As PBH binaries form in the early Universe, they emit gravitational waves continuously, slowly shrink, and ultimately merge. Nevertheless, some of them would disrupt, due to tidal forces of surrounding PBHs, before the completion of the merger phase \cite{2018PhRvD..98b3536K, 2019JCAP...02..018R}. In fact, the orbital parameters of the formed binaries determine their merger time. Given that PBHs are randomly distributed in the Universe, the orbital parameters for each formed binary will be unique. Therefore, some binaries have already merged, some of them merge in the present-time Universe, and some will merge in the future. Thus, today's merger rate will be enormously increased by those PBH binaries expected to merge in the present-time Universe. Therefore, the LIGO-Virgo observations can be justified by the fact that PBHs must constitute a very small fraction of dark matter in the mentioned mechanism of PBH binary formation \cite{2020PhRvD.102l3524H, 2021JCAP...03..068H, 2022PhLB..82937040C, 2022PhRvD.106l3526F}. However, it should be noted that both mechanisms are still valid and their predictions of the contribution of PBHs to dark matter are still being validated through gravitational wave detectors.

The total merger event rate per unit time and volume can be specified by convolving the merger rate of PBHs in every single dark matter halo with the halo mass function
\begin{equation}\label{tot_mer}
\mathcal{R}=\int_{M_{\rm c}}\frac{dn}{dM_{\rm vir}} \Gamma(M_{\rm vir})dM_{\rm vir},
\end{equation}
where $M_{\rm vir}$ is the halo virial mass, and $M_{\rm c}\simeq 400 M_{\odot}$ is the minimum mass of dark matter halos whose signals can reach the present-time Universe while containing PBHs with the mass of $M_{\rm PBH}=30M_{\odot}$ (see \cite{Fakhry:2020plg, Fakhry:2021tzk, Fakhry:2022uun, Fakhry:2022zum} for more details). Due to the exponential decreasing term in the halo mass function, the upper limit of integration is not significant in determining the merger rate of PBHs. On the other hand, in low-mass halos, dark matter density is expected to be higher than that in high-mass halos, which is consistent with the hierarchical dynamics of halo formation. Accordingly, the lower limit of integration plays a crucial role in the current analysis. To calculate the merger rate of PBHs within the context of the current analysis, we consider the fraction of PBHs as $f_{\rm PBH}=1$.

In Fig.~\ref{fig1}, we have indicated the merger rate of PBHs per unit time and volume for the DP mass function and compared it with that obtained for the ST mass function \cite{Fakhry:2020plg} while taking into account Einasto and NFW density profiles. To perform these calculations, relevant concentration-mass-redshift relations obtained in Ref.~\cite{okoli}, hereafter the Okoli-Afshordi concentration-mass-redshift relations, have been used. According to this figure, the merger rate of PBHs has an enhancement in smaller halos when considering the DP mass function as compared to the ST mass function. Meanwhile, in larger halos, the merger rate of PBHs for the DP mass function is lower than the corresponding result obtained from the ST mass function. Based on the analysis concerning the importance of the minimum mass ends in the PBH scenario, it can be inferred that subhalos can potentially play a decisive role in the merger rate of PBHs. Therefore, it can be concluded that considering the high-precision DP mass function may lead to the strengthening of the merger rate of PBHs.

\begin{figure}[t!]
\begin{minipage}{1\linewidth}
\includegraphics[width=0.9\textwidth]{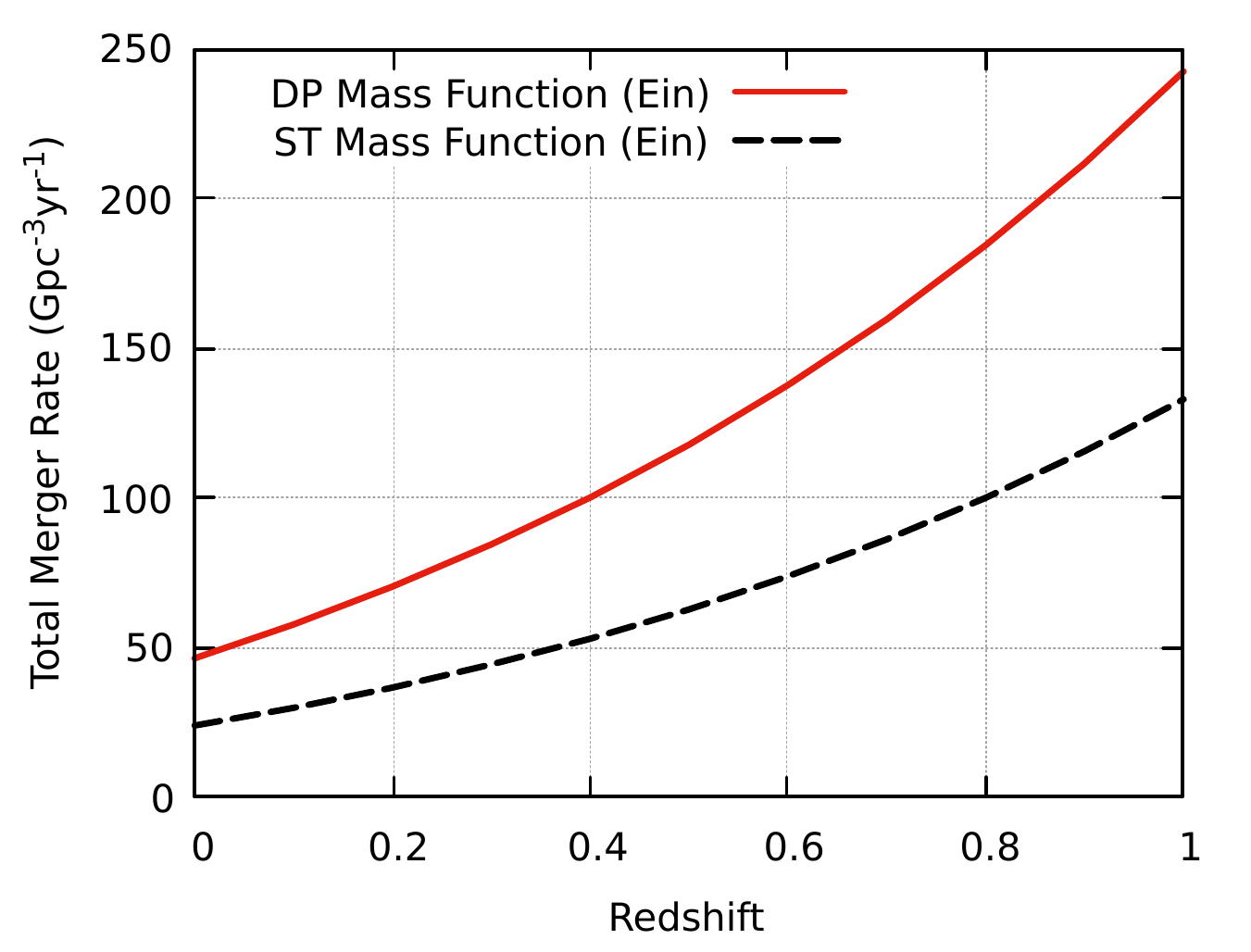}\\
\includegraphics[width=0.9\textwidth]{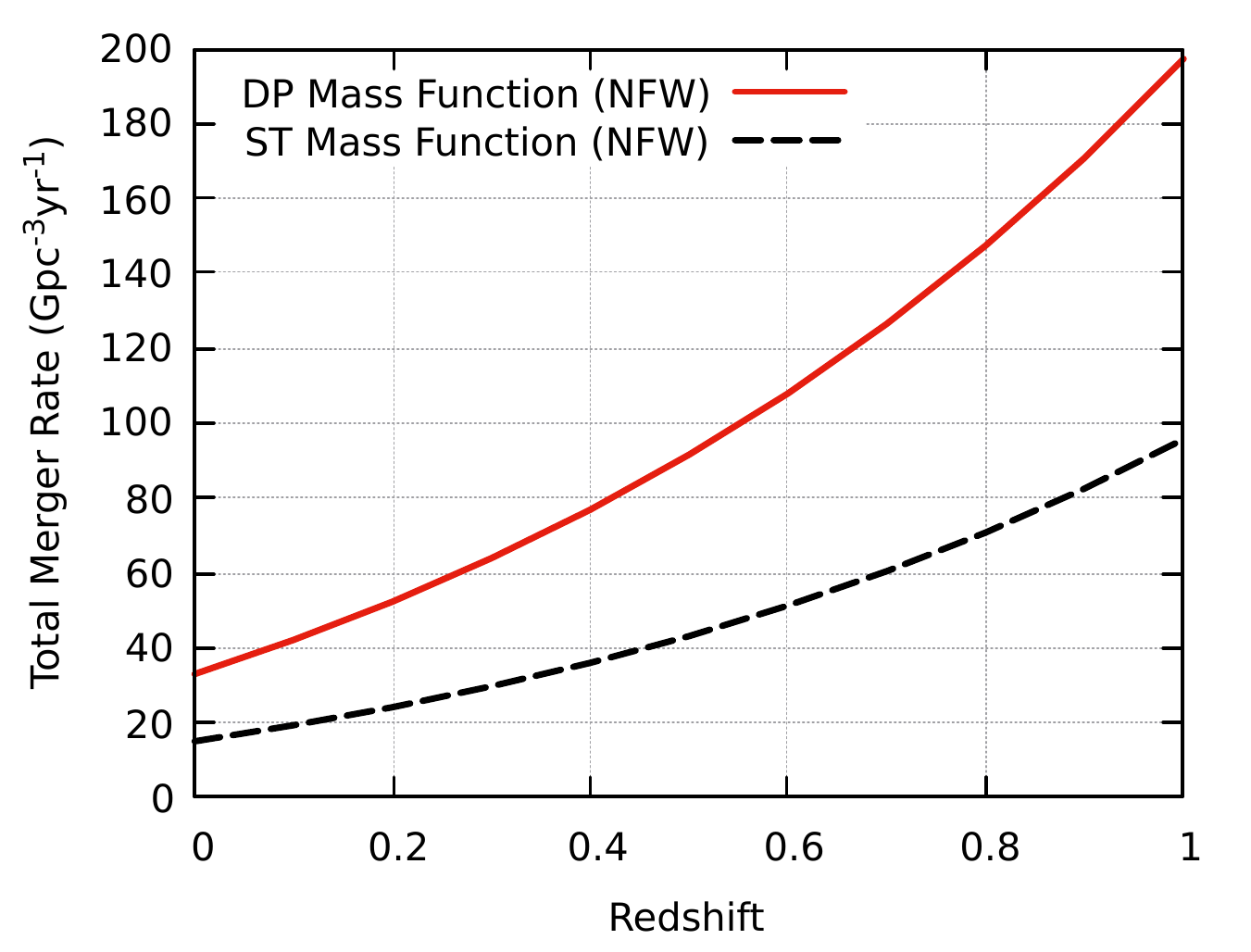}
\caption{The evolution of the total merger rate of BHs as a function of redshift for Einasto (top) and NFW (bottom) density profiles. Solid (red) lines exhibit this relation for the DP mass function, while dashed (black) lines indicate it for ST mass function.}
\label{fig3}
\end{minipage}
\end{figure}

\begin{figure}[t!]
\begin{minipage}{1\linewidth}
\includegraphics[width=0.92\textwidth]{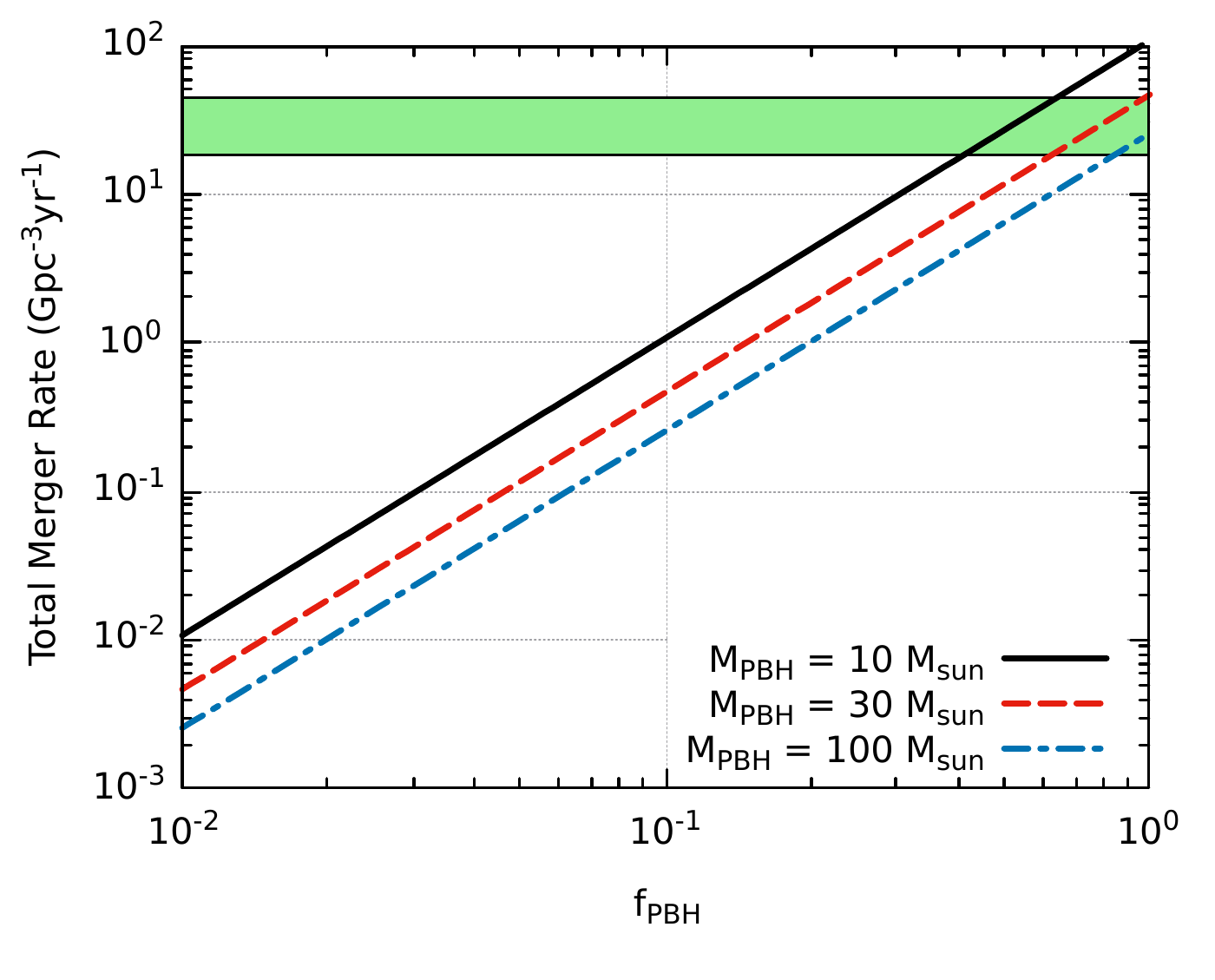}
\caption{Total merger event rate of PBHs while considering the DP mass function with respect to the PBH fraction and mass. The solid (black), dashed (red), and dot-dashed (blue) lines demonstrate this relation for a PBH mass of $M_{\rm PBH} = 10, 30$, and $100 M_{\odot}$, respectively. The shaded (green) band represents the total merger rate of PBHs estimated by the LIGO-Virgo detectors during the latest observing run, i.e., $(17.9\mbox{-}44)\,{\rm Gpc^{-3}yr^{-1}}$.}
\label{fig4}
\end{minipage}
\end{figure}

In Fig.~\ref{fig2}, we have quantified the relative enhancement of the merger rate of PBHs in terms of halo mass for DP and ST mass functions in the present-time Universe. As can be seen from this figure, the relative enhancement of the merger rate of PBHs by considering the DP mass function occurs for halo masses less than $\sim 10^{12}M_{\odot}$, while a relative reduction of the merger rate is observed for masses greater than $\sim 10^{12}M_{\odot}$. Even though in the cumulative analysis, the relative reduction of the merger rate of PBHs in cluster-sized halos is negligible, this is likely due to the repulsive effect of the cosmological constant, which opposes accretion and merging processes at a large scale. Accordingly, as expected, the physical effects that directly affect the shape of the mass function lead to the strengthening of the merger rate of PBHs. In addition, assuming the PBH mass to be $30M_{\odot}$, the amplification of the PBH merger rate when taking into account the DP mass function is in the range of $(106 \pm 13)\%$.

One of the most interesting topics to study has always been the evolution of the merger rate of BHs. It is also possible to detect GWs at higher redshifts as the precision of relevant instruments can be improved. Currently, the LIGO-Virgo detectors are capable of detecting binary mergers up to $z \simeq 0.75$, which approximately corresponds to a comoving volume of around $50 \,{\rm Gpc^{3}}$. Regarding this, through the halo mass function and the concentration parameter, Eq.~\eqref{tot_mer} is dependent on redshift. In Fig.~\ref{fig3}, we have depicted the redshift evolution of the merger event rate of PBHs for NFW and the Einasto density profiles, wherein the relevant results of DP and ST mass functions have been compared. It is evident that changes in redshift directly affect the merger rate of PBHs. This is because halo merger trees and hierarchical dynamics could have led to more subhalos at higher redshifts. As a result, PBH merger rates in the past were higher than those in the present-time Universe. Moreover, it can be concluded that the merger rate of PBHs for the DP mass function is higher than the corresponding results obtained from the ST mass function in the whole range of the late-time Universe. This is where the importance of a high-precision mass function comes into play because it could potentially provide a more accurate prediction of the accumulation of PBH mergers in the late-time Universe. 

\begin{table*}[t] 
\caption{Total merger rate of PBHs for different PBH masses, i.e., $M_{\rm pbh} = 10,20, 30, 50,$ and $100\,M_{\odot}$, considering the ellipsoidal-collapse halo models in terms of the NFW and the Einasto density profiles, at the present-time Universe $(z=0)$. Amplification percentages and the merger rate of PBHs have been provided while taking into account the DP and ST mass functions.}
\centering
\begin{tabular}{c|c|c|c|c}
\hline
\hline
PBH Mass $(M_{\odot})$ & Density Profile &$\mathcal{R}_{\rm DP} (\rm Gpc^{-3}\rm yr^{-1})$&$\mathcal{R}_{\rm ST}(\rm Gpc^{-3}\rm yr^{-1})$ & Amplification Percentage\\ [0.5ex]
\hline
10 & NFW & $79.46$ & $33.41$ & $137\%$\\
10 & Einasto & $107.5$ & $51.25$ & $109\%$\\
\hline
20 & NFW & $49.60$ & $22.38$ & $121\%$\\
20 & Einasto & $68.58$ & $35.01$ & $96\%$\\
\hline
30 & NFW & $33.01$ & $15.06$ & $119\%$\\
30 & Einasto & $46.51$ & $24.03$ & $93\%$\\
\hline
50 & NFW & $25.01$ & $11.49$ & $117\%$\\
50 & Einasto & $35.70$ & $18.60$ & $91\%$\\
\hline
100 & NFW & $17.64$ & $8.18$ & $115\%$\\
100 & Einasto & $25.60$ & $13.48$ & $89\%$\\
\hline
\hline
\end{tabular}
\label{table1}
\end{table*}

So far, our analysis was carried out assuming the mass of the PBHs to be $30 M_{\odot}$, and their contribution to the dark matter to be maximum. However, it is interesting to calculate the merger rate of PBHs according to their fractions and masses. In Fig.~\ref{fig4}, we have illustrated the merger rate of PBHs for the DP mass function in terms of the upper bounds on the fraction of PBHs and their masses. In this calculation, the Einasto density profile is considered. Moreover, the shaded band indicates the total merger rate of BHs estimated by the LIGO-Virgo detectors during the latest observing run, i.e., $(17.9\mbox{-}44)\,{\rm Gpc^{-3}yr^{-1}}$ \cite{LIGOScientific:2021djp}. It can be observed that the merger rate of PBHs is inversely proportional to their masses but directly proportional to their fractions. This is because the number density of PBHs changes inversely with their masses. In addition, despite the theoretical uncertainties, the merger rate of PBHs for the DP mass function will fall within the LIGO-Virgo sensitivity if $f_{\rm PBH}\gtrsim \mathcal{O}(10^{-1})$. Meanwhile, compared to the corresponding results derived from the ST mass function, the present analysis can potentially improve the constraints on the abundance of PBHs.

In Table \ref{table1}, we have provided the relevant results of the merger rate of PBHs in terms of various PBH masses, NFW and Einasto density profiles, and DP and ST mass functions. Furthermore, we have presented the relative amplification of the merger rate of PBHs while considering the DP mass function. Specifically, in comparison with our previous study \cite{Fakhry:2020plg}, the results indicate that considering the DP mass function can lead to a relative strengthening of the merger rate of stellar-mass PBHs. Additionally, the results exhibit that the relative enhancement of the merger rate of PBHs changes inversely with their masses. For example, in the mass range of $M_{\rm PBH} = (10\mbox{-}100)M_{\odot}$, the enhancement of the merger rate of PBH with the mass of $10M_{\odot}$ is in the range of $(123\pm 14)\%$, while the corresponding result for PBHs with the mass of $100M_{\odot}$ takes place in the range of $(102\pm 13)\%$.
%%%%%%%%%%%%%%%%%%%%%%%%%%%%%%%%%%%%%%%%%%%%%%%%%%%%%%%%
\section{Conclusions} \label{sec:iv}
In the standard model of cosmology, also known as the $\Lambda$CDM model, dark matter halos are fundamental units of cosmological structures that form and evolve hierarchically based on the cosmological perturbation theory. Observational data show that dark matter halos have been distributed randomly throughout the Universe. Therefore, they can be analyzed from a statistical point of view. In the classification of dark matter halos, special attention should be paid to their local quantities. To this end, the halo mass function plays a crucial role. Hence, it is of fundamental importance to determine a high-precision halo mass function. Using numerical simulations, several attempts have been carried out to determine the exact shape of the halo mass function, which have led to appropriate predictions. However, the black-box nature of numerical simulations makes it difficult to distinguish the role of physical mechanisms influencing the shaping of the halo mass function. Based on this, semianalytical approaches can more accurately demarcate physical effects in explaining the statistical properties of dark matter halos.

In this work, we investigate the effect of a high-precision semianalytical mass function on the merger rate of PBHs in dark matter halos. For this purpose, we have initially introduced the relevant theoretical framework for dark matter halo models. Dark matter distribution in galactic halos can be well described by the halo density profile. Another crucial factor that represents the dark matter density in the central region of halos is known as the concentration-mass-redshift relation. In this work, we have used NFW and Einasto density profiles as well as Okoli-Afshordi concentration-mass-redshift relations. Furthermore, we have introduced the semianalytical DP mass function, which makes use of tidal torque, dynamical friction, and the cosmological constant to shape the halo mass function and increase its precision.

Relying on the PBH scenario and in the framework of an ellipsoidal-collapse dark matter halo model, we have calculated the merger rate of PBHs considering the DP mass function and compared it qualitatively and quantitatively with the relevant results of the ST mass function. Our findings demonstrate that, compared to that obtained for the ST mass function, the merger rate of PBHs with the DP mass function experiences a relative amplification for halos smaller than $10^{12}M_{\odot}$, while it has a relative reduction for those halos grater than $10^{12}M_{\odot}$. Given that smaller halos have a prominent contribution to the merger rate of PBHs, it can be argued that using the DP mass function leads to the amplification of the merger rate of PBHs. In this regard, taking the mass of PBHs as $M_{\rm PBH}=30M_{\odot}$, the present analysis estimates the amplification of the merger rate to be in the range of $(106 \pm 13)\%$. One has to mention that in these calculations, Poissonian fluctuations caused by PBHs play a crucial role in strengthening the power spectrum of dark matter on small scales, and naturally strengthening the merger rate of PBHs, which is not considered in the previous work associated with the ST mass function.

We have also calculated the redshift evolution of the merger event rate of PBHs for the DP mass function and compared it with the corresponding result obtained from the ST mass functions. The results exhibit that the merger rate of PBHs for the DP mass function is higher than that derived from the ST mass function during the late-time Universe. This is due to the importance of a high-precision mass function that can provide more accurate predictions of PBH mergers.

Finally, we have calculated the merger rate of PBHs for the DP mass function as a function of their mass and fraction and compared the results with the BH mergers estimated by the LIGO-Virgo detectors during the latest observing run, i.e., $(17.9\mbox{-}44)\,{\rm Gpc^{-3}yr^{-1}}$. Our findings indicate that the merger rate of PBHs is inversely proportional to their mass but directly proportional to their fraction. Based on this argument, the results show that the merger rate of PBHs can be in the range of the LIGO-Virgo sensitivity if $f_{\rm PBH} \gtrsim \mathcal{O}(10^{-1})$. On the other hand, compared to the results obtained for the ST mass function, it is deduced that the constraints on the fraction of PBHs can be strengthened while considering the DP mass function. Moreover, our results show that the relative amplification of the merger rate of PBHs changes inversely with their masses.

Although the current analysis improves our previous results of the merger rate of PBHs in dark matter halos, the conditions considered therein are not necessarily satisfied. Hence, the analysis presented in this work may include theoretical uncertainties. For example, possible three-body encounters as more complicated channels for the binary formation, and binary disruption by tidal forces of surrounding compact objects. It is hoped that these uncertainties can be minimized in the near future.


\begin{thebibliography}{99} 

\bibitem{Mandel:2021smh}
I. Mandel and F.S. Broekgaarden,
``Rates of compact object coalescences",
{\it Living Rev. Rel.} {\bf 25}, 1 (2022).

\bibitem{Olsen:2022pin}
S. Olsen {\it et al.},
``New binary black hole mergers in the LIGO-Virgo O3a data'',
{\it Phys. Rev. D} \textbf{106}, 043009 (2022).

\bibitem{Abbott:2016blz}
B.P. Abbott {\it et al.},
``Observation of gravitational waves from a binary black hole merger'',
{\it Phys. Rev. Lett.} {\bf 116}, 061102 (2016).

\bibitem{Abbott:2016nmj}
B.P. Abbott {\it et al.},
``GW151226: Observation of gravitational waves from a 22-solar-mass binary black hole coalescence'',
{\it Phys. Rev. Lett.} {\bf 116}, 241103 (2016).

\bibitem{TheLIGOScientific:2016src}
B.P. Abbott {\it et al.},
``Tests of general relativity with GW150914'',
{\it Phys. Rev. Lett.} {\bf 116}, 221101 (2016); Erratum:~{\it Phys. Rev. Lett.} {\bf 121}, 129902 (2018).

\bibitem{Abbott:2020khf}
R. Abbott {\it et al.},
``GW190814: Gravitational waves from the coalescence of a 23 solar mass black hole with a 2.6 solar mass compact object'',
{\it Astrophys. J. Lett.} {\bf 896}, L44 (2020).

\bibitem{Abbott:2020tfl}
R. Abbott {\it et al.},
``GW190521: A binary black hole merger with a total mass of $150~\rm M_{\odot}$'',
{\it Phys. Rev. Lett.} {\bf 125}, 101102 (2020).

\bibitem{fishbach2021}
M. Fishbach {\it et al.},
``When are LIGO/Virgo's big black hole mergers?",
{\it Astrophys. J.} {\bf 912}, 98 (2021).

\bibitem{rodriguez2021}
C.L. Rodriguez {\it et al.},
``The observed rate of binary black hole mergers can be entirely explained by globular clusters",
{\it Res. Notes AAS} {\bf 5}, 19 (2021).

\bibitem{bird2016}
S. Bird {\it et al.}, 
``Did Ligo detect dark matter?",
{\it Phys. Rev. Lett.} {\bf 116}, 201301 (2016).

\bibitem{Clesse:2016vqa}
S. Clesse and J. Garc\'\i{}a-Bellido,
``The clustering of massive primordial black holes as dark matter: Measuring their mass distribution with advanced LIGO",
{\it Phys. Dark Univ.} \textbf{15}, 142 (2017).

\bibitem{sasaki2016}
M. Sasaki, T. Suyama, T. Tanaka and S. Yokoyama,
``Primordial black hole scenario for the gravitational-wave event GW150914",
{\it Phys. Rev. Lett.} {\bf 117}, 061101 (2016).

\bibitem{Spergel:1999mh}
D.N. Spergel and P.J. Steinhardt,
``Observational evidence for selfinteracting cold dark matter,''
{\it Phys. Rev. Lett.} \textbf{84}, 3760 (2000).

\bibitem{MACHO:2000qbb}
C. Alcock \textit{et al.},
``The MACHO project: Microlensing results from 5.7 years of LMC observations,''
{\it Astrophys. J.} \textbf{542}, 281 (2000).

\bibitem{Liebling:2012fv}
S.L. Liebling and C. Palenzuela,
``Dynamical boson stars",
{\it Living Rev. Rel.} \textbf{15}, 6 (2012).

\bibitem{Roszkowski:2017nbc}
L. Roszkowski, E.M. Sessolo and S. Trojanowski,
``WIMP dark matter candidates and searches-current status and future prospects",
{\it Rept. Prog. Phys.} \textbf{81}, 066201 (2018).

\bibitem{ADMX:2019uok}
T. Braine \textit{et al.},
``Extended search for the invisible axion with the axion dark matter experiment",
{\it Phys. Rev. Lett.} \textbf{124}, 101303 (2020).

\bibitem{DiGiovanni:2020frc}
F. Di Giovanni, S. Fakhry, N. Sanchis-Gual, J.C. Degollado and J.A. Font,
``Dynamical formation and stability of fermion-boson stars",
{\it Phys. Rev. D} \textbf{102}, 084063 (2020).

\bibitem{DiGiovanni:2021vlu}
F.Di Giovanni, S. Fakhry, N. Sanchis-Gual, J.C. Degollado and J.A. Font,
``A stabilization mechanism for excited fermion-boson stars",
{\it Class. Quant. Grav.} \textbf{38}, 194001 (2021).

\bibitem{zeldovic1967}
Y.B. Zel'dovich and I.D. Novikov,
``The hypothesis of cores retarded during expansion and the hot cosmological model",
{\it Soviet Astron. AJ} (Engl. Transl.) {\it 10}, 602 (1967).

\bibitem{Hawking:1971ei}
S. Hawking,
``Gravitationally collapsed objects of very low mass",
{\it Mon. Not. Roy. Astron. Soc.} \textbf{152}, 75 (1971).

\bibitem{Carr:1974nx}
B.J. Carr and S. Hawking,
``Black holes in the early universe",
{\it Mon. Not. Roy. Astron. Soc.} \textbf{168}, 399 (1974).

\bibitem{niemar1999}
J.C. Niemeyer and K. Jedamzik,
``Dynamics of primordial black hole formation",
{\it Phys. Rev. D} {\bf 59}, 124013 (1999).

\bibitem{shibati1999}
M. Shibata and M. Sasaki,
``Black hole formation in the Friedmann universe: Formulation and computation in numerical relativity",
{\it Phys. Rev. D} {\bf 60}, 084002 (1999).

\bibitem{plonarov2007}
A.G. Polnarev and I. Musco,
``Curvature profiles as initial conditions for primordial black hole formation",
{\it Class. Quant. Grav.} {\bf 24}, 1405 (2007).

\bibitem{musco2013}
I. Musco and J.C. Miller,
``Primordial black hole formation in the early universe: Critical behaviour and selfsimilarity",
{\it Class. Quant. Grav.} {\bf 30}, 145009 (2013).

\bibitem{young2014}
S. Young, C.T. Byrnes and M. Sasaki,
``Calculating the mass fraction of primordial black holes",
{\it J. Cosmol. Astropart. Phys.} {\bf 1407}, 045 (2014).

\bibitem{bloomfield15}
J. Bloomfield, D. Bulhosa and S. Face,
``Formalism for primordial black hole formation in spherical symmetry",
arXiv:1504.02071.

\bibitem{allahyari2017}
A. Allahyari, J.T. Firouzjaee and A.A. Abolhasani,
``Primordial black holes in linear and non-linear regimes",
{\it J. Cosmol. Astropart. Phys.} {\bf 1706}, 041 (2017).

\bibitem{Sasaki:2018dmp}
M. Sasaki, T. Suyama, T. Tanaka and S. Yokoyama,
``Primordial black holes-perspectives in gravitational wave astronomy",
{\it Class. Quant. Grav.} \textbf{35}, 063001 (2018).

\bibitem{LIGOScientific:2018mvr}
B.P. Abbott \textit{et al.},
``GWTC-1: A gravitational-wave transient catalog of compact binary mergers observed by LIGO and Virgo during the first and second observing runs",
{\it Phys. Rev. X} {\bf 9}, 031040 (2019).

\bibitem{LIGOScientific:2020ibl}
R. Abbott \textit{et al.},
``GWTC-2: Compact binary coalescences observed by LIGO and Virgo during the first half of the third observing run",
{\it Phys. Rev. X} {\bf 11}, 021053 (2021).

\bibitem{LIGOScientific:2021djp}
R. Abbott \textit{et al.},
``GWTC-3: Compact binary coalescences observed by LIGO and Virgo during the second part of the third observing run",
arXiv:2111.03606.

\bibitem{Fakhry:2020plg}
S. Fakhry, J.T. Firouzjaee and M. Farhoudi,
``Primordial black hole merger rate in ellipsoidal-collapse dark matter halo models",
{\it Phys. Rev. D} \textbf{103}, 123014 (2021).

\bibitem{Fakhry:2021tzk}
S. Fakhry, M. Naseri, J.T. Firouzjaee and M. Farhoudi,
``Primordial black hole merger rate in self-interacting dark matter halo models",
{\it Phys. Rev. D} \textbf{105}, 043525 (2022).

\bibitem{Fakhry:2022uun}
S. Fakhry, Z. Salehnia, A. Shirmohammadi and J.T. Firouzjaee,
``The merger rate of primordial black hole-neutron star binaries in ellipsoidal-collapse dark matter halo models",
{\it Astrophys. J.} \textbf{941}, 36 (2022).

\bibitem{Fakhry:2022zum}
S. Fakhry, S.S. Tabasi and J.T. Firouzjaee,
``On the merger rate of primordial black holes in cosmic voids",
arXiv:2210.13558.

\bibitem{2011PhRvL.107b1302S} 
B.D. Sherwin {\it et al.},
``Evidence for dark energy from the cosmic microwave background alone using the atacama cosmology telescope lensing measurements",
{\it Phys. Rev. Lett.} \textbf{107}, 021302 (2011).

\bibitem{Yang:2014okp}
W. Yang and L. Xu,
``Testing coupled dark energy with large scale structure observation",
{\it J. Cosmol. Astropart. Phys.} \textbf{08}, 034 (2014).

\bibitem{Huterer:2017buf}
D. Huterer and D.L. Shafer,
``Dark energy two decades after: Observables, probes, consistency tests",
{\it Rept. Prog. Phys.} \textbf{81}, 016901 (2018).

\bibitem{DiValentino:2019ffd}
E. Di Valentino, A. Melchiorri, O. Mena and S. Vagnozzi,
``Interacting dark energy in the early 2020s: A promising solution to the $H_0$ and cosmic shear tensions",
{\it Phys. Dark Univ.} \textbf{30}, 100666 (2020).

\bibitem{Y:2021ybx}
M.G. Yengejeh, A. Behnamfard, S. Fakhry and J.T. Firouzjaee,
``The integrated Sachs-Wolfe effect in 4D Einstein-Gauss-Bonnet gravity",
{\it Phys. Dark Univ.} \textbf{35}, 100918 (2022).

\bibitem{Yengejeh:2022tpa}
M.G. Yengejeh, S. Fakhry, J.T. Firouzjaee and H. Fathi,
``The integrated Sachs-Wolfe effect in interacting dark matter-dark energy models",
{\it Phys. Dark Univ.} \textbf{39}, 101144 (2023).

\bibitem{ps1974}
W.H. Press and P. Schechter,
``Formation of galaxies and clusters of galaxies by self-similar gravitational condensation",
{\it Astrophys. J.} {\bf 187}, 425 (1974).

\bibitem{Sheth:1999mn}
R.K. Sheth and G. Tormen,
``Large scale bias and the peak background split",
{\it Mon. Not. Roy. Astron. Soc.} \textbf{308}, 119 (1999).

\bibitem{Sheth:1999su}
R.K. Sheth, H.J. Mo and G. Tormen,
``Ellipsoidal collapse and an improved model for the number and spatial distribution of dark matter haloes",
{\it Mon. Not. Roy. Astron. Soc.} \textbf{323}, 1 (2001).

\bibitem{Sheth:2002}
R.K. Sheth and G. Tormen,
``An excursion set model of hierarchical clustering: ellipsoidal collapse and the moving barrier",
{\it Mon. Not. Roy. Astron. Soc.} \textbf{329}, 61 (2002).

\bibitem{Reed:2003sq}
D.~Reed {\it et al.},
``Evolution of the mass function of dark matter haloes",
{\it Mon. Not. Roy. Astron. Soc.} \textbf{346}, 565 (2003).

\bibitem{Reed:2003hp}
D. Reed {\it et al.},
``Evolution of the density profiles of dark matter halos'',
{\it Mon. Not. Roy. Astron. Soc.} \textbf{357}, 82 (2005).

\bibitem{ein} 
J. Einasto,
``On the construction of a composite model for the galaxy and on the determination of the system of galactic parameters",
{\it Trudy Astrofizicheskogo Instituta Alma-Ata} \textbf{5}, 87 (1965).

\bibitem{Jaffe:1983iv}
W. Jaffe,
``A Simple model for the distribution of light in spherical galaxies'',
{\it Mon. Not. Roy. Astron. Soc.} \textbf{202}, 995 (1983).

\bibitem{deZeeuw:1985sk}
T. de Zeeuw,
``Elliptical galaxies with separable potentials'',
{\it Mon. Not. Roy. Astron. Soc.} \textbf{216}, 273 (1985).

\bibitem{Hernquist:1990be}
L. Hernquist,
``An analytical model for spherical galaxies and bulges'',
{\it Astrophys. J.} \textbf{356}, 359 (1990).

\bibitem{dehnen}
W. Dehnen,
``A family of potential-density pairs for spherical galaxies and bulges",
{\it Mon. Not. Roy. Astron. Soc.} \textbf{265}, 250 (1993).

\bibitem{Navarro:1995iw}
J.F. Navarro, C.S. Frenk and S.D.M. White,
``The structure of cold dark matter halos'',
{\it Astrophys. J.} \textbf{462}, 563 (1996).

\bibitem{prada}
F. Prada {\it et al.},
``Halo concentrations in the standard $\Lambda$ cold dark matter cosmology",
{\it Mon. Not. Roy. Astron. Soc.} \textbf{423}, 3018 (2012).

\bibitem{dutton}
A.A. Dutton and A.V. Macci\`o,
``Cold dark matter haloes in the Planck era: Evolution of structural parameters for Einasto and NFW profiles",
{\it Mon. Not. Roy. Astron. Soc.} \textbf{441}, 3359 (2014).

\bibitem{okoli}
C. Okoli and N. Afshordi,
``Concentration, ellipsoidal collapse, and the densest dark matter haloes",
{\it Mon. Not. Roy. Astron. Soc.} \textbf{456}, 3068 (2016).

\bibitem{ludlow}
A.D. Ludlow {\it et al.},
``The mass-concentration-redshift relation of cold and warm dark matter haloes",
{\it Mon. Not. Roy. Astron. Soc.} \textbf{460}, 1214 (2016).

\bibitem{WMAP:2003elm} 
D.N. Spergel \textit{et al.}, 
``First year Wilkinson Microwave Anisotropy Probe (WMAP) observations: Determination of cosmological parameters",
{\it Astrophys. J. Suppl.} \textbf{148}, 175 (2003).

\bibitem{DelPopolo:2007dna}
A. Del Popolo,
``Dark matter and structure formation a review",
{\it Astron. Rep.} \textbf{51}, 169 (2007).

\bibitem{2011ApJS..192...18K}
E. Komatsu \textit{et al.},
``Seven-year Wilkinson Microwave Anisotropy Probe (WMAP) observations: Cosmological interpretation",
{\it Astrophys. J. Suppl.} \textbf{192}, 18 (2011).

\bibitem{delpopolo2013}
A. Del Popolo,
``Non-baryonic dark matter in cosmology",
{\it AIP Conf. Proc.} \textbf{1548}, 2 (2013).

\bibitem{2013ApJ...779...86S} 
K.T. Story {\it et al.},
``A Measurement of the cosmic microwave background damping tail from the 2500-square-degree SPT-SZ survey",
{\it Astrophys. J.} \textbf{779}, 86 (2013). 

\bibitem{Das:2013zf}
S. Das \textit{et al.} 
``The atacama cosmology telescope: Temperature and gravitational lensing power spectrum measurements from three seasons of data",
{\it J. Cosmol. Astropart. Phys.} \textbf{04}, 014 (2014).

\bibitem{DelPopolo:2013qba}
A. Del Popolo,
``Nonbaryonic dark matter in cosmology",
{\it Int. J. Mod. Phys. D} \textbf{23}, 1430005 (2014).

\bibitem{Planck:2015fie}
P.A.R. Ade \textit{et al.},
``Planck 2015 results. XIII. cosmological parameters",
{\it Astron. Astrophys.} \textbf{594}, A13 (2016).

\bibitem{Hiotelis:2005jh}
N. Hiotelis and A. Del Popolo,
``On the reliability of merger-trees and the mass growth histories of dark matter haloes",
{\it Astrophys. Space Sci.} \textbf{301}, 167 (2006).

\bibitem{Hiotelisapp}
N. Hiotelis and A. Del Popolo,
``Anomalous diffusion models for the formation of dark matter haloes",
{\it Mon. Not. Roy. Astron. Soc.} \textbf{436}, 163 (2013). 

\bibitem{Efstathiou:1988tk}
G. Efstathiou, C.S. Frenk, S.D.M. White and M. Davis,
``Gravitational clustering from scale free initial conditions",
{\it Mon. Not. Roy. Astron. Soc.} \textbf{235}, 715 (1988).

\bibitem{Gross:1997wv}
M.A.K. Gross, R.S. Somerville, J.R. Primack, J. Holtzman and A. Klypin,
``CDM-variant cosmological models 1.: Simulations and preliminary comparisons",
{\it Mon. Not. Roy. Astron. Soc.} \textbf{301}, 81 (1998).

\bibitem{Jenkins:2000bv}
A. Jenkins {\it et al.},
``The Mass function of dark matter halos",
{\it Mon. Not. Roy. Astron. Soc.} \textbf{321}, 372 (2001).

\bibitem{White:2002at}
M.J. White,
``The mass function",
{\it Astrophys. J. Suppl.} \textbf{143}, 241 (2002).

\bibitem{Lukic:2007fc}
Z. Lukic, K. Heitmann, S. Habib, S. Bashinsky and P.M. Ricker,
``The halo mass function: High redshift evolution and universality",
{\it Astrophys. J.} \textbf{671}, 1160 (2007).

\bibitem{Reed:2006rw}
D. Reed, R. Bower, C. Frenk, A. Jenkins and T. Theuns,
``The halo mass function from the dark ages through the present day",
{\it Mon. Not. Roy. Astron. Soc.} \textbf{374}, 2 (2007).

\bibitem{2011ApJ...740..102K}
A.A. Klypin, S. Trujillo-Gomez and J. Primack,
``Dark matter halos in the standard cosmological model: Results from the Bolshoi simulation",
{\it Astrophys. J.} \textbf{740}, 102 (2011).

\bibitem{1991ApJ...379..440B}
J.R. Bond, S. Cole, G. Efstathiou and N. Kaiser,
``Excursion set mass functions for hierarchical Gaussian fluctuations,"
{\it Astrophys. J.} \textbf{379}, 440 (1991).

\bibitem{1993MNRAS.262..627L}
C. Lacey and S. Cole,
``Merger rates in hierarchical models of galaxy formation",
{\it Mon. Not. Roy. Astron. Soc.} \textbf{262}, 627 (1993).

\bibitem{DelPopolo:2006gn}
A. Del Popolo and S.I. Yesilyurt,
``On the cosmological mass function theory",
{\it Astron. Rep.} \textbf{51}, 709 (2007).

\bibitem{Popolo:1998fz}
A. Del Popolo and M. Gambera,
``Tidal torques and the clusters of galaxies evolution",
{\it Astron. Astrophys.} \textbf{337}, 96 (1998).

\bibitem{DelPopolo:2006bk}
A. Del Popolo,
``On the average comoving number density of halos",
{\it Astrophys. J.} \textbf{637}, 12 (2006).

\bibitem{DelPopolo:2006shv}
A. Del Popolo,
``Some improvements to the spherical collapse model",
{\it Astron. Astrophys.} \textbf{454}, 17 (2006).

\bibitem{DelPopolo:2017snx}
A. Del Popolo, F. Pace and M. Le Delliou,
``A high precision semi-analytic mass function",
{\it J. Cosmol. Astropart. Phys.} \textbf{03}, 032 (2017).

\bibitem{1994ApJ...427...72A}
 V. Antonuccio-Delogu and S. Colafrancesco,
``Dynamical friction, secondary infall, and the evolution of clusters of galaxies",
{\it Astrophys. J.} \textbf{427}, 72 (1994).

\bibitem{Colafrancesco:1994ne}
S. Colafrancesco, V. Antonuccio-Delogu and A. Del Popolo,
``On the dynamical origin of bias in clusters of galaxies",
{\it Astrophys. J.} \textbf{455}, 32 (1995).

\bibitem{2009ApJ...698.2093D}
A. Del Popolo,
``The cusp/core problem and the secondary infall model",
{\it Astrophys. J.} \textbf{698}, 2093 (2009).

\bibitem{2009A&A...502..733D}
A. Del Popolo and P. Kroupa,
``Density profiles of dark matter haloes on galactic and cluster scales",
{\it Astron. Astrophys.} \textbf{502}, 733 (2009).

\bibitem{2006ApJ...646..881W} 
M.S. Warren, K. Abazajian, D.E. Holz and L. Teodoro,
``Precision determination of the mass function of dark matter halos",
{\it Astrophys. J.} \textbf{646}, 881 (2006).

\bibitem{2007MNRAS.374....2R}
D.S. Reed {\it et al.},
``The halo mass function from the dark ages through the present day",
{\it Mon. Not. Roy. Astron. Soc.} \textbf{374}, 2 (2007).

\bibitem{2010MNRAS.403.1353C} 
M. Crocce, P. Fosalba, F.J. Castander and E. Gazta{\~n}aga, 
``Simulating the Universe with MICE: The abundance of massive clusters",
{\it Mon. Not. Roy. Astron. Soc.} \textbf{403}, 1353 (2010).

\bibitem{2011MNRAS.410.1911C} 
L. Courtin {\it et al.}.
``Imprints of dark energy on cosmic structure formation - II. Non-universality of the halo mass function",
{\it Mon. Not. Roy. Astron. Soc.} \textbf{410}, 1911 (2011).

\bibitem{2011ApJ...732..122B} 
S. Bhattacharya {\it et al.},
``Mass function predictions beyond {\ensuremath{\Lambda}}CDM",
{\it Astrophys. J.} \textbf{732}, 122 (2011).

\bibitem{2012MNRAS.426.2046A} 
R.E. Angulo {\it et al.},
``Scaling relations for galaxy clusters in the Millennium-XXL simulation",
{\it Mon. Not. Roy. Astron. Soc.} \textbf{426}, 2046 (2011).

\bibitem{Afshordi:2003zb}
N. Afshordi, P. McDonald and D.N. Spergel,
``Primordial black holes as dark matter: The power spectrum and evaporation of early structures",
{\it Astrophys. J. Lett.} \textbf{594}, L71 (2003).

\bibitem{Kashlinsky:2016sdv}
A. Kashlinsky,
``LIGO gravitational wave detection, primordial black holes and the near-IR cosmic infrared background anisotropies",
{\it Astrophys. J. Lett.} \textbf{823}, L25 (2016)/

\bibitem{OLeary:2008myb}
R.M. O'Leary, B. Kocsis and A. Loeb,
``Gravitational waves from scattering of stellar-mass black holes in galactic nuclei",
{\it Mon. Not. Roy. Astron. Soc.} \textbf{395}, 2127 (2009).

\bibitem{Quinlan1989}
G.D. Quinlan and S.L. Shapiro,
``Dynamical evolution of dense clusters of compact stars",
{\it Astrophys. J.} \textbf{343}, 725 (1989).

\bibitem{Carr:2018rid}
B. Carr and J. Silk,
``Primordial black holes as generators of cosmic structures",
{\it Mon. Not. Roy. Astron. Soc.} \textbf{478}, 3756 (2018).

\bibitem{Korol:2019jud}
V. Korol, I. Mandel, M.C. Miller, R.P. Church and M.B. Davies,
``Merger rates in primordial black hole clusters without initial binaries",
{\it Mon. Not. Roy. Astron. Soc.} \textbf{496}, 994 (2020).

\bibitem{Trashorras:2020mwn}
M. Trashorras, J. Garc\'\i{}a-Bellido and S. Nesseris,
``The clustering dynamics of primordial black boles in $N$-body simulations",
{\it Universe} \textbf{7}, 18 (2021).

\bibitem{Franciolini:2022ewd}
G. Franciolini, K. Kritos, E. Berti and J. Silk,
``Primordial black hole mergers from three-body interactions",
{\it Phys. Rev. D} \textbf{106}, 083529 (2022).

\bibitem{peters}
P.C. Peters,
``Gravitational radiation and the motion of two point masses",
{\it Phys. Rev.} {\bf 136}, B1224 (1964).

\bibitem{Mouri2002}
H. Mouri and Y. Taniguchi,
``Runaway merging of black holes: Analytical constraint on the timescale",
{\it Astrophys. J. Lett.} \textbf{566}, L17 (2002).

\bibitem{prada2012}
F. Prada {\it et al.},
``Halo concentrations in the standard $\Lambda$ cold dark matter cosmology",
{\it Mon. Not. Roy. Astron. Soc.} \textbf{423}, 3018 (2012).

\bibitem{2017PhRvD..96l3523A} 
Y. Ali-Ha{\"\i}moud, E.D. Kovetz and M. Kamionkowski,
``Merger rate of primordial black-hole binaries", 
{\it Phys. Rev. D} \textbf{96}, 123523 (2017).

\bibitem{2020PhRvD.101d3015V}
V. Vaskonen and H. Veerm{\"a}e, 
``Lower bound on the primordial black hole merger rate",
{\it Phys. Rev. D} \textbf{101}, 043015 (2020).

\bibitem{2018PhRvD..98b3536K} 
 B.J. Kavanagh, D. Gaggero and G. Bertone,
``Merger rate of a subdominant population of primordial black holes",
{\it Phys. Rev. D} \textbf{98}, 023536 (2018).

\bibitem{2019JCAP...02..018R} 
M. Raidal, C. Spethmann, V. Vaskonen and H. Veerm{\"a}e,
``Formation and evolution of primordial black hole binaries in the early universe",
{J. Cosmol. Astropart. Phys.} \textbf{2019}, 18 (2019). 

\bibitem{2020PhRvD.102l3524H}
A. Hall, A.D. Gow, C.T. Byrnes,
``Bayesian analysis of LIGO-Virgo mergers: Primordial versus astrophysical black hole populations",
{\it Phys. Rev. D} \textbf{102}, 123524 (2020).

\bibitem{2021JCAP...03..068H}
G. H{\"u}tsi, M. Raidal, V. Vaskonen and H. Veerm{\"a}e,
``Two populations of LIGO-Virgo black holes",
{J. Cosmol. Astropart. Phys.} \textbf{2021}, 68 (2021).

\bibitem{2022PhLB..82937040C} 
Z.C. Chen, C. Yuan and Q.G. Huang,
``Confronting the primordial black hole scenario with the gravitational-wave events detected by LIGO-Virgo", {Phys. Lett. B} \textbf{829}, 137040 (2022).

\bibitem{2022PhRvD.106l3526F}
G. Franciolini, I. Musco, P. Pani and A. Urbano
``From inflation to black hole mergers and back again: Gravitational-wave data-driven constraints on inflationary scenarios with a first-principle model of primordial black holes across the QCD epoch",
{\it Phys. Rev. D} \textbf{106}, 123526 (2022).




\end{thebibliography}
\end{document}